\documentclass[structabstract]{aa}  
\usepackage{graphicx}
\usepackage{txfonts}
\begin{document}
   \title{Modelling the light-curve of KIC 12557548b: an extrasolar 
planet with a comet like tail}


   \author{J. Budaj
          }

   \institute{Astronomical Institute, Slovak Academy of Sciences,
              05960 Tatranska Lomnica, Slovak Republic,
              \email{budaj@ta3.sk}
             }

   \date{Received ; accepted }

 
  \abstract
{An object with a very peculiar light-curve was discovered 
recently using Kepler data. 
Authors argue that this object may be a transiting disintegrating 
planet with a comet like dusty tail. 
Light-curves of some eclipsing binaries  
may have features analogous to this light-curve and it is very
interesting to see whether they are also caused by the same effects 
and put them in a more general context.
}
{The aim of the present paper is to verify the model suggested by 
the discoverers by the light-curve modelling and put constraints on 
the geometry of the dust region and various dust properties.}
{We modify the code SHELLSPEC designed for modelling of the interacting 
binaries to calculate the light-curves of stars with such planets.
We take into account the Mie absorption and scattering on spherical 
dust grains of various sizes assuming realistic dust opacities and 
phase functions and finite radius of the source of the scattered light.
}
{
The planet light-curve is reanalysed using long and short cadence
Kepler observations from the first 14 quarters.
Orbital period of the planet was improved. 
We prove that the peculiar light-curve of this objects
is in agreement with the idea of a planet with a comet like tail. 
Light-curve has a prominent pre-transit brightening and a less 
prominent post-transit brightening. Both are caused by the forward 
scattering and are a strong function of the particle size.
This feature enabled us to estimate a typical particle size (radius) 
in the dust tail of about 0.1-1 micron.
However, there is an indication that the particle size changes 
along the tail. Larger particles better reproduce the pre-transit 
brightening and transit core while smaller particles are more 
compatible with the egress and post-transit brightening.
Dust density in the tail is a steep decreasing function of the distance
from the planet which indicates a significant tail destruction caused 
by the star. We also argue that the 'planet' does not 
show uniform behaviour but may have at least two constituents.
This light-curve with pre-transit brightening is analogous to
the light-curve of $\epsilon$ Aur with mid-eclipse brightening and 
forward scattering plays a significant role in such eclipsing systems. 
}
{}

\maketitle
%

\section{Introduction}

The exoplanet candidate KIC012557548b has been discovered 
recently by Rappaport et al. (\cite{rappaport12}). It was discovered 
from Kepler long cadence data (Borucki et al. \cite{borucki11}) 
obtained during first two quarters. This exoplanet is very unique. 
Unlike all other exoplanets it exhibits strong variability in the 
transit depth. For some period of time transits even disappear. 
Shape of the transit is highly asymmetric with a significant 
brightening just before the eclipse, sharp ingress followed by a 
smooth egress. Planet has also extremely short period of
0.65356(1) days (15.6854 hours). Rappaport et al. (\cite{rappaport12})
suggested that the planet has size not larger than Mercury and is 
slowly disintegrating/evaporating what creates a comet like tail made 
of pyroxene grains.
Perez-Becker \& Chiang \cite{perez13} constructed a 
radiative-hydrodynamic model of the atmospheric escape from such
low mass rocky planets.
The hypothesis that a close-in planet can have a cometary-like 
tail was first suggested by Schneider et al. (\cite{schneider98})
and revisited by Mura et al. (\cite{mura11}).
The transit light-curve of dusty comets was first investigated by
Lecavelier des Etangs et al. (\cite{lecavelier99}).  
There is another class of objects which may look very different but 
may have features analogous to this light-curve. 
$\epsilon$ Aur is an interacting binary with 
the longest known orbital period, 27.1 yr.
The primary star, which is the main source of light, may be either
a young massive F0Ia super-giant or an evolved post-AGB star 
(see Guinan et al. \cite{guinan12}, Hoard et al. \cite{hoard} 
and references therein). 
The star is partially eclipsed by a dark dusty disk and the light-curve
shows a very unusual shallow mid-eclipse brightening (MEB). 
It was suggested that the disk has a central hole and is inclined out 
of the orbital plane so that the star can peek through the hole 
(Carroll et al. \cite{carroll}). Recently, Budaj (\cite{budaj11a}) 
proposed that MEB might be due to the flared disk 
geometry and forward scattering on dust and that one does not 
necessarily have to see the primary star through the hole in the disk 
during the eclipse. Calculation of Muthumariappan \& Parthasarathy 
(\cite{muthumariappan12}) indeed indicate that the disk is flared and 
the hole is present but not seen at such almost edge-on inclination.
Shallow MEB was observed also in some other long period eclipsing
binaries, for example in AZ Cas (Galan et al. \cite{galan12}).

Planetary transits are usually modelled using the analytical 
formulae of Mandel \& Agol (\cite{mandel02}). This approach assumes 
spherical shape of the objects. JKTEBOP code can calculate and solve 
the transits numerically assuming the shape of bi-axial 
ellipsoids (Southworth \cite{southworth12}).
BEER algorithm (Faigler \& Mazeh \cite{faigler11}) calculates 
analytical light-curves including Doppler boosting.
The EVILMC code developed by Jackson et al. (\cite{jackson12})
can model transit light-curves assuming the Roche shape.
The SHELLSPEC code of Budaj \& Richards (\cite{budaj04})
can calculate planetary light-curves assuming the Roche
shape, the new model of the reflection effect, and the 
circum-stellar/planetary material.
Modelling the transit light-curve of KIC012557548, however, will 
be different and will require at least some modifications 
to these codes or a new approach.
Shortly before the submission of this manuscript the light-curve
of this planet was modelled independently by 
Brogi et al. (\cite{brogi12}). These authors started with thorough
data reduction of raw Kepler long cadence photometry from first six 
quarters. They assumed 1D model of the dust cloud in which the vertical 
dimension of the cloud was negligible compared with the stellar 
radius. They also modelled dust extinction as a free parameter
and assumed analytical Henyey-Geenstein phase functions 
and point source approximation for the scattered light. 
The particle size was estimated mainly from the overall shape of 
the transit.

In this paper we first revisit the light-curve and orbital 
period using long as well as short cadence Kepler observations 
from the first 14 quadratures (Sections \ref{obs}, \ref{tail}). 
Then we modify the code SHELLSPEC to model light-curves of such objects. 
In Section \ref{dust} we will calculate real opacities and 
phase functions of pyroxene and other similar dust grains. 
In Section \ref{s4} we construct a 3D model 
of the dust cloud, calculate the radiative transfer along the line 
of sight and take into account a finite dimension of the source of 
light. We will estimate the particle size from the pre-transit 
brightening feature which is most sensitive to this parameter.
This will enable us to verify whether the shape of the light-curve 
is in agreement with the idea of a planet with the comet like tail, 
put constraints on the particle size and geometry of the dust region,
and put this interesting object which may look like a comet,
behave like an eclipsing interacting binary, but be an exoplanet,
into a more general context.

\begin{figure}
\centering
\includegraphics[angle=0,width=8.3cm]{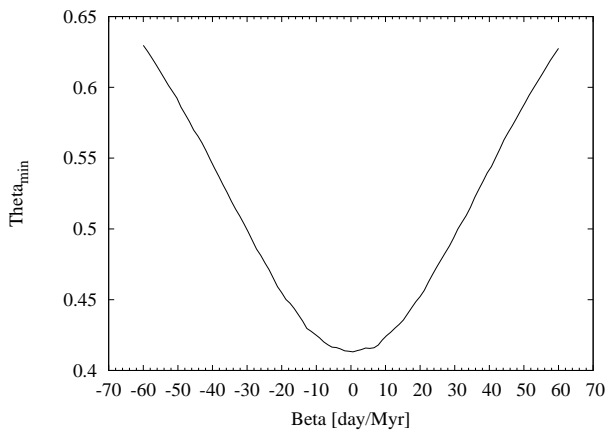}
\caption{
The search for a long term period change.
The minimum of the theta parameter corresponds to the period change 
rate $\beta=0.3\pm0.5$ days/Myr which means that there is no 
significant long term period variability.
}
\label{f1}
\end{figure}

\section{Observations}
\label{obs}

In this paper we use the publicly available Kepler data from the first
14 quarters (Borucki et al. \cite{borucki11}) in the form of 
SAP flux.\footnote{We avoided using the PDCSAP flux since astrophysical
signal could be artificially removed from the raw data during
the automated, not optimized photometry decorrelation.} 
These are long cadence observations with the exposure time 
of about 30 minutes and short cadence observations with the exposure 
time of about 1 minute. Each quarter has a different flux level. 
In our first 
step we scaled the flux from all quarters to the same level.

The long cadence data were used to verify and estimate 
the orbital period of the planet and search for other periods 
in these data. We used two independent methods.
The first was the Fourier method. A rough scan over the data was 
made to find an approximate orbital period.
Consequently, we phased the data with this preliminary period and set 
the phase zero point such that the transits occur approximately at 
the phase of 0.5. Then we split the data into many chunks each covering 
one orbital period, fit the straight line into each chunk of data
covering 
phases 0-1 and remove (divide out) the linear trend from the data.
During this process, a phase interval of $<0.42,0.58>$, which contains
the transit, was excluded from the fitting.
In this way we removed from the light-curve any changes on time scales 
longer than the orbital period. 
The advantage of this particular form of detrending is that it
does not introduce any additional non-linear trend into the phased
light-curve data. 
Then we again searched for the periods 
and found the final value of the frequency 1.530100(4) cyc/day
which corresponds to 0.6535521(15) days (or 15.68525 hours).
We estimated the error by performing Monte Carlo simulations generating
and analysing about 100 artificial datasets.
This value of the period is in very good agreement with the value 
reported in the discovery paper (0.65356(1) days), however, the error 
of our analysis comprising much more data is about 7 times 
smaller.

\begin{figure*}
\centerline{
\includegraphics[angle=-90,width=8.8cm]{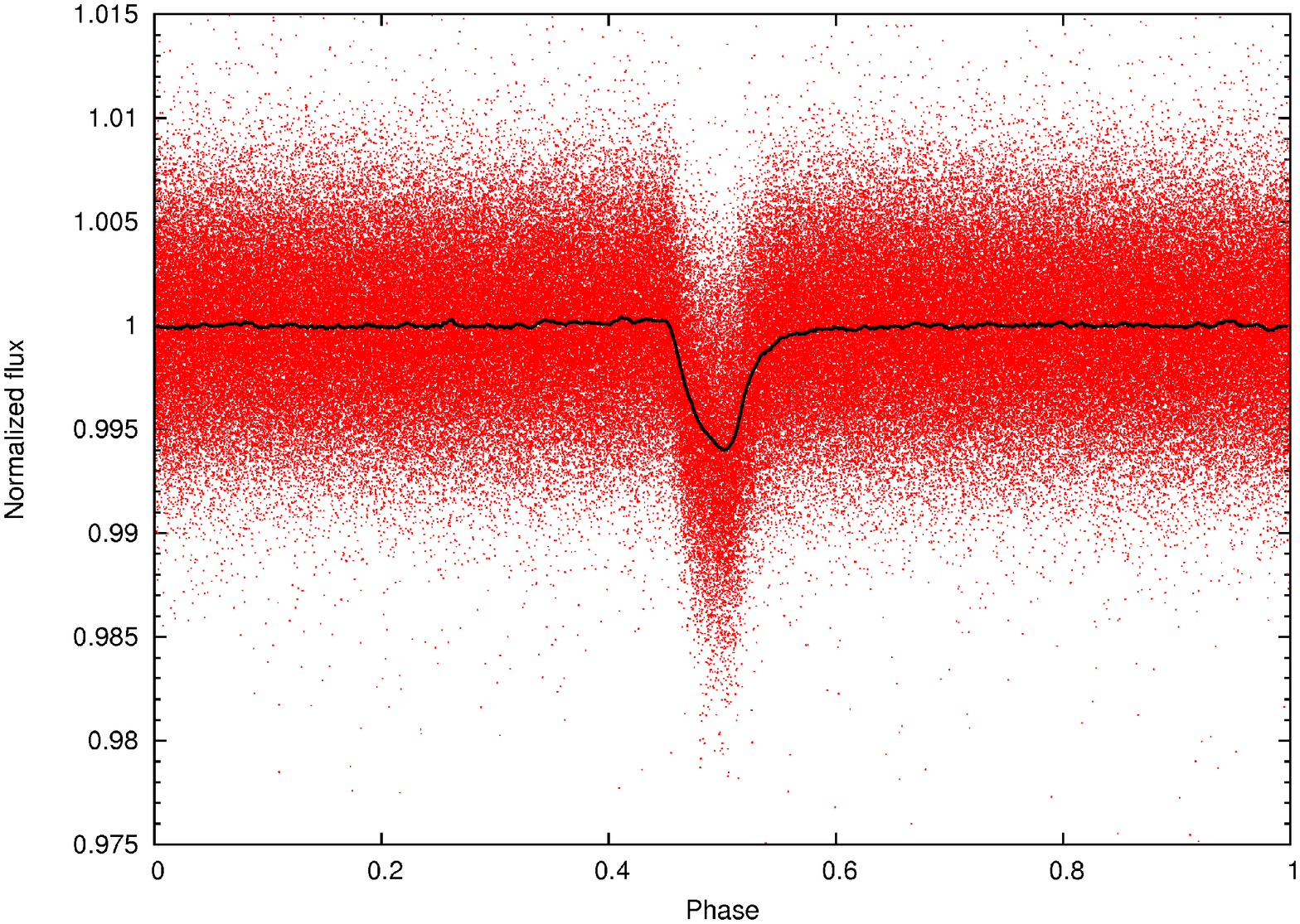}
\includegraphics[angle=-90,width=8.8cm]{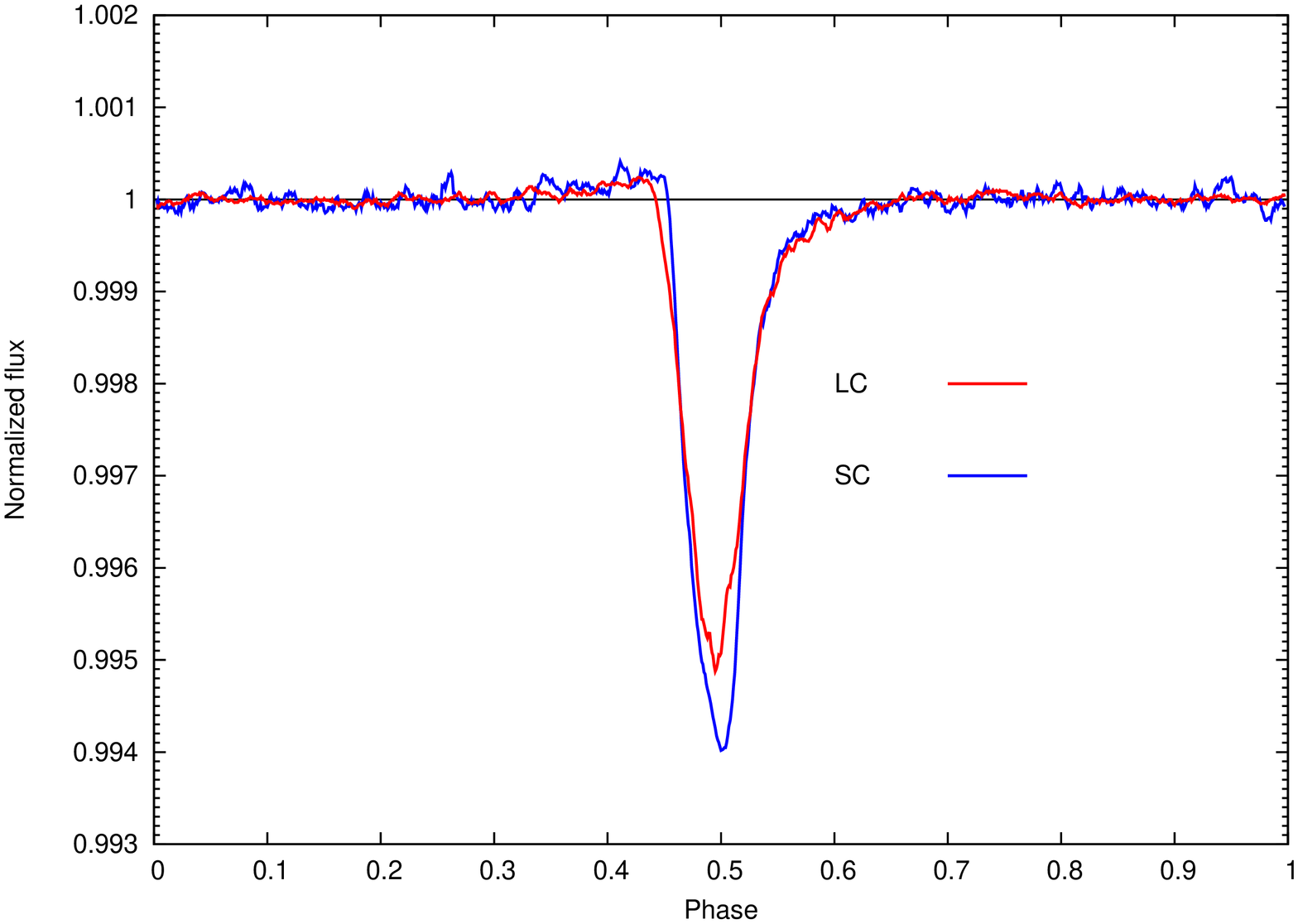}
}
\caption{
{\it Left}: Short cadence Kepler observations (dots) phased with 
the orbital period. The linear trend was subtracted and data were 
smoothed (line) with the running window with the width of 0.01.
{\it Right:} A comparison between the short cadence (SC, blue) and 
the long cadence (LC, red) Kepler transit light-curves.
}
\label{distribsc}
\end{figure*}

The light-curve obtained with this procedure and folded with this 
improved period was then subject to a running window averaging.
We used a running window with the width of 0.01 and the step of 0.001 
in the phase units and calculated the averaged (phase, flux) points 
within each window position.
Notice that the width of the window corresponds to about 0.16 hours
and its smoothing effect on the data is negligible in comparison with 
the exposure time or finite angular dimension of the star.
This light-curve was then used for the modelling.
It is depicted in the Figure \ref{distribsc}
and described in more detail in the Section \ref{s4}. 
The standard deviation of points of this binned light-curve with 
phases 0.9-0.1 which are not spoiled by the dusty tail is only 
$3\times10^{-5}$ which is about half of that mentioned by 
Brogi et al. (\cite{brogi12}). It illustrates that our method works 
very well and that there are no serious discontinuities between 
the individual epochs.

Phase dispersion minimization method (PDM, Stellingwerf 
\cite{stellingwerf78}) was used next. This method is very convenient 
in cases with non-sinusoidal phase variability and with non-continuous 
sampling. The method minimizes the variance of the data with respect 
to the mean light-curve which is described by the parameter theta. 
We used the latest 100 bin version (pdm2b4.13).
Period search revealed that the most significant variability has 
a period of about 22.7 days. 
This period and its amplitude are not strict but consist of 
several similar periods which pop up in different data subsets.
It might be associated with the stellar differential rotation and 
some spots at different latitudes. 
This variability makes the identification of other periods more 
complicated. The method confirms the orbital period found by 
the Fourier method.

Since the planet is very close to the parent star it is apparently
being disintegrated and loosing material. One might expect all kinds
of interaction that could lead to the long term period evolution.
That is why we also searched for possible long term 
changes of the orbital period, $P$, with the PDM method. We assumed 
that the period changes linearly with time $t$ according to the 
following expression
\begin{equation}
P=P_{0}+\beta t
\end{equation}
where $P_{0}$ is the period obtained without the period change and 
$\beta$ is the period change rate in d/Myr. We did not find
any significant long term period changes during the time span of 
the Kepler observations.
This is illustrated in the Figure \ref{f1} which displays the 
dependence of theta as a function of $\beta$.
We fitted a parabola to this curve and obtained
$\beta=0.3\pm0.5$ days/Myr which means that there is no significant
evidence for the long term period change. The error was estimated 
by means of Monte Carlo simulations. We generated and analysed 
100 artificial data sets 
with the same standard deviation as the original data.

The short cadence Kepler observations were treated using the 
same method. We extracted the SAP flux, normalized it, divided into 
individual epochs, fitted and subtracted the linear trend and smoothed
with the running window which was 0.01 phase units wide with the
step of 0.001. The result is plotted in the Figure \ref{distribsc}.
The comparison between the long and short cadence light-curves
is depicted in the Figure \ref{distribsc}. 
The short cadence
light-curve is slightly deeper and has slightly more pronounced
pre-transit brightening feature.

\section{Evolution of the tail}
\label{tail}

\begin{figure*}
\centerline{
\includegraphics[angle=-90,width=8.8cm]{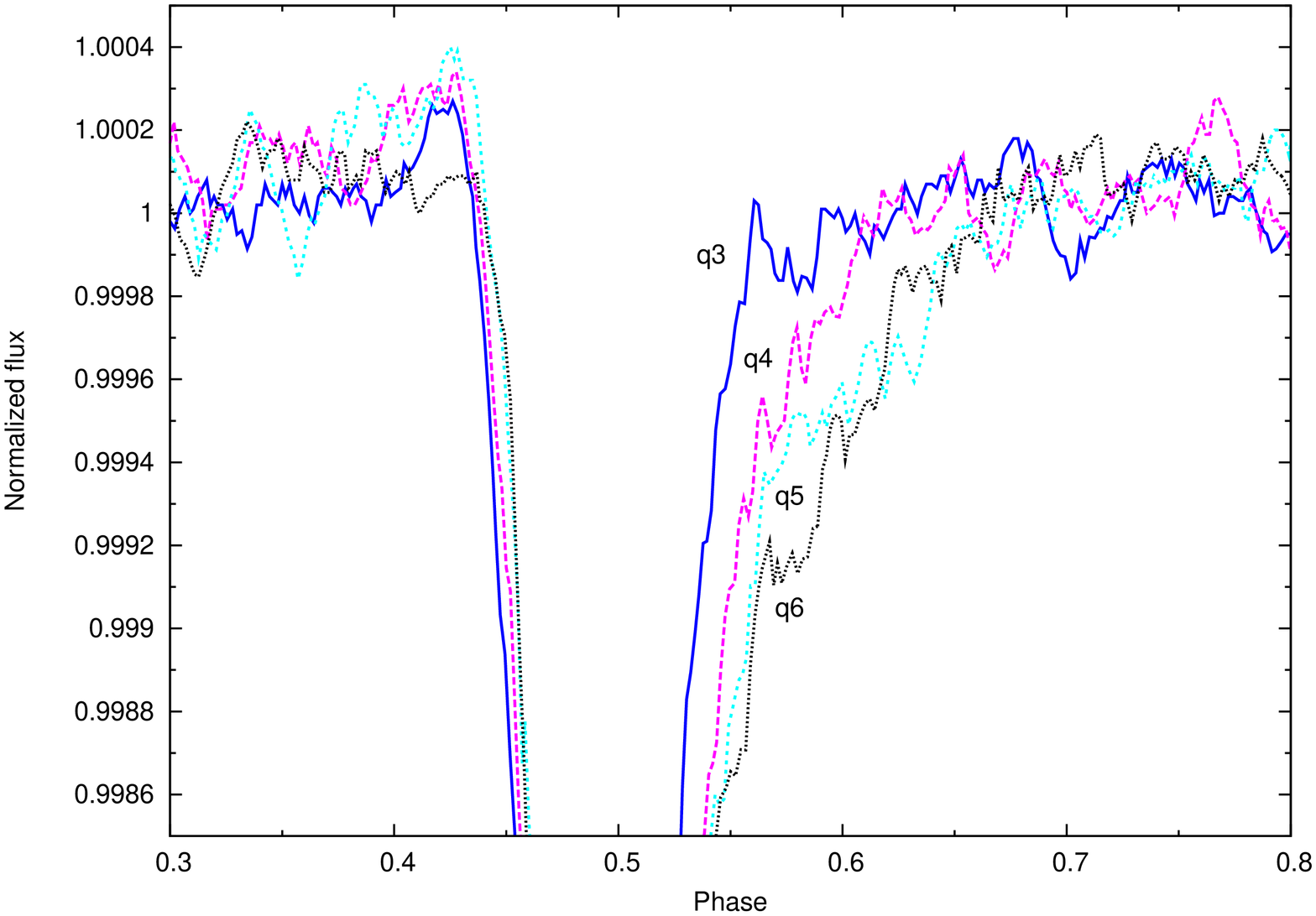}
\includegraphics[angle=-90,width=8.8cm]{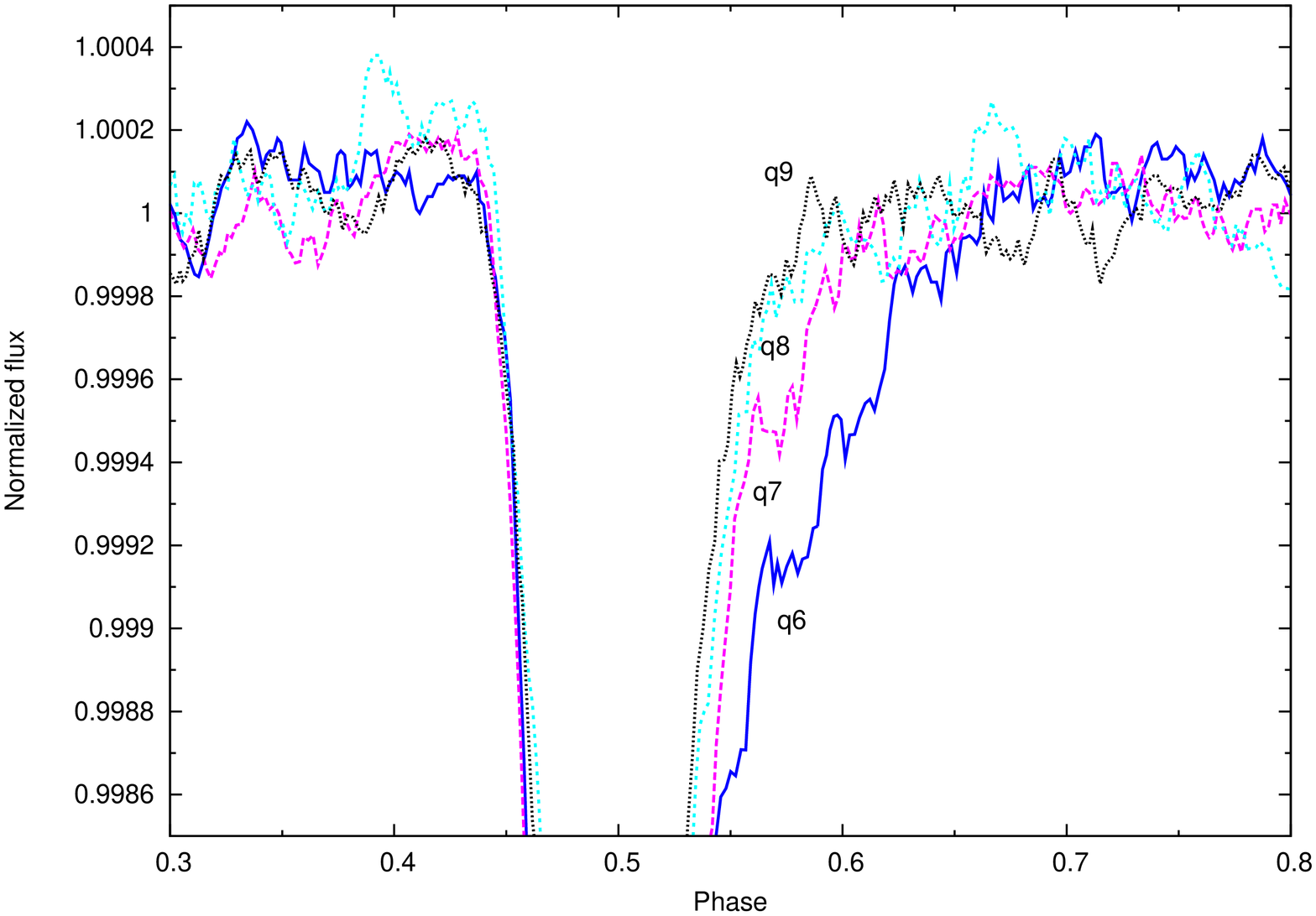}
}
\caption{
Evolution of the comet-like tail.
{\it Left}: There is a  progressive strengthening of the absorption in 
the tail between Kepler quarters 3 to 6.
{\it Right:} There is a progressive diminishing of the absorption in
the tail between Kepler quarters 6 to 9.
}
\label{tail1}
\end{figure*}

Apart from the known variability in the transit depth 
which can easily be seen in the rough data 
(Rappaport et al. \cite{rappaport12}) there may also be
changes in the 'cometary' tail. These changes are not that easily 
detected since the tail is much finer and not as deep as 
the core of the transit.
To verify this idea we constructed an averaged light-curve for 
the long cadence observations during each Kepler quarter with the same 
method as described in the previous section. In this case, 
a twice as wide box-car window was used, compared to the previous case. 
These light-curves were then inspected for variability.
One can indeed see in 
the Figure \ref{tail1} that there are apparent changes 
in the tail on the timescale of about 1 yr. Namely, the tail gets 
more pronounced during quarters 3 to 6, then diminishes during 
quarters 6 to 9. The story continues and the tail progressively gets 
stronger during quarters 9 to 11, and again ceases during 
quarters 11 to 13 but this is not shown in these pictures.
This indicates that the changes may be periodic with a period of about
one and a half year.
Curiously, these changes in the tail on these timescales and
in these averaged light-curves can be even stronger than the changes
in the transit core which happened during the quarters 3-6!
The changes in the tail do not seem to correlate with the changes 
in the transit core.

The origin of this variability needs to be explored.
One can speculate that it may be associated with the magnetic activity
of the star or planet.
One cannot also exclude that these cycles may originate from 
the variability in the transit core which reflects the variability
of the dense material in the close vicinity of the planet which
feeds the tail. In either case and in a more general context this 
variability represents a new type of star-planet interaction 
(Shkolnik et al. \cite{shkolnik08}).
Note that these changes in the tail might also cause spurious changes 
in the orbital period due to their asymmetric nature.
It indicates that the planet is not uniform
and has at least two components: a sort of 'coma' responsible for 
the transit core and the tail responsible for the egress.

\section{Optical properties of the dust}
\label{dust}

\begin{figure}
\centering
\includegraphics[angle=-90,width=8.8cm]{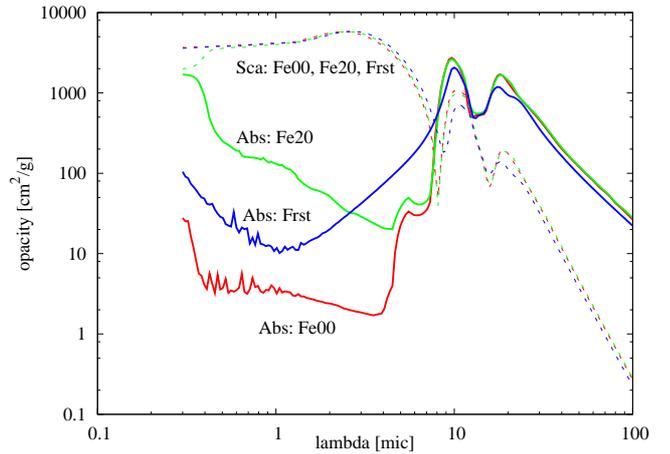}
\caption{
Opacities of forsterite and pyroxene for different
iron content. Blue (Frst) - forsterite, red (Fe00) - pyroxene
with 0\% of iron (enstatite), green (Fe20)- pyroxene with 20\% of iron.
Solid line - absorption, dashed line - scattering. Calculated
assuming the characteristic radius of the grains 1 micron.
}
\label{f2a}
\end{figure}

\begin{figure}
\centering
\includegraphics[angle=-90,width=8.8cm]{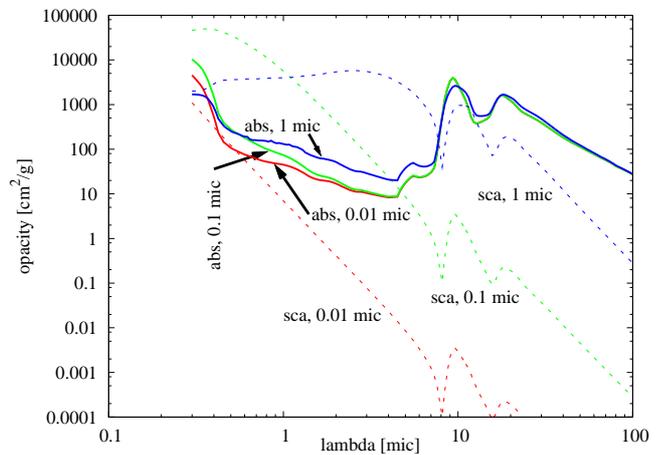}
\caption{
Opacities of pyroxene with 20\% of iron as a function of particle 
radius. Red - 0.01 micron, green - 0.1 micron, blue - 1 micron.
Solid line - absorption, dashed line - scattering.
}
\label{f2b}
\end{figure}

\begin{figure}
\centering
\includegraphics[angle=-90,width=8.8cm]{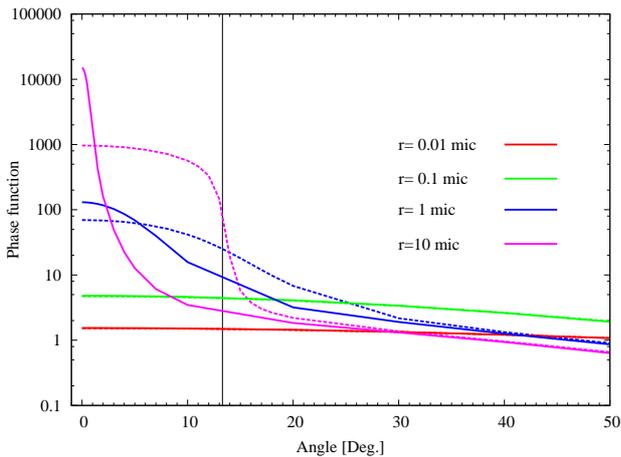}
\caption{
Phase functions at 650 nm for different particle radii. 
Full line - phase functions assuming a point source of light.
Dashed line -phase functions which take into account the finite 
dimension of the stellar disk. Notice the strong forward scattering 
peak for larger particles.
Vertical line illustrates the angular dimension of the stellar disk 
as seen from the planet.
}
\label{f2}
\end{figure}

According to Rappaport et al. (\cite{rappaport12}) the comet like tail 
of the planet may consist mainly of dust made of pyroxene grains.
Optical properties of the pyroxene and another silicate
of the olivine family - forsterite were calculated using Mie theory
and Mie scattering code of Kocifaj (\cite{kocifaj04}) and      
Kocifaj et al. (\cite{kocifaj08}) which is based on Bohren \& Huffman
(\cite{bohren83}). By optical properties we mean opacities for 
scattering and absorption and phase functions.
The complex index of refraction 
of pyroxene was taken from Dorschner et al. (\cite{dorschner95}).
The complex index of refraction 
of forsterite was taken from J\"ager et al.(\cite{jager}).
We assumed spherical particles of different sizes.
To suppress the ripple structure that would appear in 
the phase function of spherical mono-disperse particles, 
the poly-disperse Deirmendjian (\cite{deirmendjian64}) distribution 
of particle sizes was assumed.

Optical properties of pyroxene ($Mg_{1-x}Fe_{x}SiO_{3}$) in the optical
region are quite sensitive to the amount of iron in the mineral. 
It is illustrated on the example of 1 micron grains in 
the Figure \ref{f2a}.
Opacity in the optical and near infrared for micron size particles
is dominated by the scattering.
Scattering is not very sensitive to the amount of iron.
However, the absorption opacity in the optical and near infrared
is very sensitive to the amount of iron and increases with higher iron
content. Its direct impact on the spectrum is thus suppressed by 
the strong scattering. On the other hand, stronger absorption of iron 
rich pyroxene will lead to enhanced heating of the grains which
might affect its sublimation and evaporation. 
Opacities of forsterite ($Mg_{2}SiO_{4}$) for 1 micron grains
are also shown in that picture. Its scattering properties are very 
similar to pyroxene but it differs in absorption in the optical
and near IR regions.
Notice, that the total (i.e. absorption plus scattering) opacity is 
almost gray in the optical and near infrared region for 1 micron grains.
The chemical composition of the dust in this system is 
not known and that is why we carried out calculations for all these
species. However, if not mentioned otherwise, 
(in the light of the above mentioned accounts) we assumed a default
model composed of the pyroxene grains with a nonzero
iron content ($x=0.2$) where 20\% of Mg atoms were replaced by Fe.

This situation may change for particles of different size.
Figure \ref{f2b} compares the opacities of 0.01, 0.1 and 1 micron 
grains of iron contaminated pyroxene.
The true absorption of smaller grains is generally slightly lower 
in the optical and NIR regions. The scattering opacity 
overcomes the absorption in the optical and NIR region
for larger grains. Notice that scattering on 1 micron grains
is rather grey in the optical and NIR region and governs optical
properties in the 1-8 micron region. Scattering on intermediate 
0.1 micron grains have steep color dependence and governs optical
properties at wavelengths shorter than 1 micron. 
Scattering on even smaller 0.01 micron grains is rather weak and 
has very strong $\lambda^{-4}$ wavelength dependence characteristic of 
the Rayleigh scattering regime. Absorption dominates the opacity
for such small particles. Absorption also dominates scattering
for wavelengths longer than 7 micron for all particle sizes smaller
than 1 micron.
Consequently, multi-wavelength transit observations in the optical 
and NIR regions might potentially constrain the particle size in this 
regime.

The phase functions for three different particle sizes at 6500 \AA\ 
are illustrated in the Figure \ref{f2} as a function of the phase 
angle. Phase angle is an angle between the original and new scattered 
beam direction.
These phase functions exhibit a strong peak near the phase angle zero
which is the so-called forward scattering peak. 
Larger particles and/or shorter wavelengths
tend to have stronger and narrower forward scattering peak than 
smaller particles and/or longer wavelengths.
Importance of this forward
scattering for the eclipsing systems during the eclipse was stressed
by Budaj (\cite{budaj11a}). 

However, these phase functions do not take into account the finite
and non-negligible angular dimension of the stellar disk as seen from
the planet which is about 26 degrees in the diameter. 
To take this effect into account one would have to split
the stellar disk into elementary surfaces and integrate over the disk.
Note that this process is mathematically analogous and similar to 
the rotational broadening of the stellar spectra. 
The difference is that while in the calculations of the rotational 
broadening the star is split into strips with constant radial velocity,
in the calculation of the scattered light, the star may be split
into arcs with constant phase angle. For phase angles greater than 
$R^{*}/a$, where $R^{*}$ is the star radius and $a$ is the semi-major 
axis, this approximation will be much more realistic than the point 
source  approximation since the arcs will resemble the strips. This 
regime involves also the pre-transit brightening 
(see Section \ref{s4}). For smaller angles it will not be that good.
Consequently, we convolve the phase functions with the 'rotational' 
broadening function (BF, Rucinski \cite{rucinski02}) of the star 
expressed in the angular units.
This function is an ellipse with the width corresponding to the angular 
dimension of the star as seen from the planet and takes into account 
the linear limb darkening of the stellar surface. 
We used the limb darkening coefficients from Claret (\cite{claret00}) 
(see the next section). 

One has to be cautious and calculate the phase functions with a very 
fine step near the zero angle because of the strong forward scattering 
and, consequently, its convolution with the BF with a very 
fine step near the edge of the broadening ellipse.
These phase functions which take into account the finite dimension
of the source of light are also depicted in the Figure \ref{f2} for
comparison. One can see that the finite dimension of the source of 
light is important for larger particles but has almost no effect for 
particles smaller than 0.1 micron. 
Broadened phase functions are re-normalized to 4$\pi$ during 
the calculations.

\section{Light-curve modelling}
\label{s4}

In this section we calculate light-curves of the exoplanet using
the code SHELLSPEC. 
This codes calculates the light-curves and spectra of interacting 
binaries or exoplanets immersed in the 3D moving circum-stellar 
(planetary) environment. It solves simple radiative transfer along 
the line of sight and the scattered light is taken into account under 
the assumption that the medium is optically thin.
It is written in Fortran77 but there is also a recent Fortran90 
version (\v{S}ejnov\'{a} et al. \cite{sejnova12}).
The code was modified so that it is possible to model dusty 
objects with comet like tails. For this purpose a few new objects
(structures) were introduced into the code and studied. They have 
the form of a cone, ring or an arc with a variable cross-section. 
Phase functions and dust opacities can be pre-calculated, loaded in 
the form of a table and interpolated to a particular wavelength 
during the execution of the program.

In these calculations the star is assumed to be a sphere with the 
radius of $R^{*}=0.65 R_{\odot}$, mass of 0.7 $M_{\odot}$ and be 
subject to the limb darkening. We used quadratic limb darkening 
coefficients for R filter from Claret (\cite{claret00}) which are 
based on the Kurucz models.
They were interpolated from the grid assuming
an effective temperature of $T_{\rm eff}=4300$K, 
surface gravity of $\log g=4.63$ (cgs), metallicity of $[M/H]=0.$, 
and micro-turbulent velocity of $v_{t}=2$ km\,s$^{-1}$.
Given the fact that the transit is sometimes missing in the 
observations, the planet itself must be rather small. 
It is its comet like tail which gives rise to the major drop in 
the light-curve. Consequently,
the planet is modelled as a 3D object in the form of a ring (or part 
of the ring) with the radius of $a=2.8 R_{\odot}$ and a non negligible
thickness (geometrical cross-section). This cross-section, $C$, 
of the ring and dust density, $\rho$, along the ring are allowed 
to change with the angle, $t$ in [rad], in the following way:
\begin {equation}
\rho(t)=\rho(0)\frac{C(0)}{C} [|t-t(0)|/\pi+1]^{A1}~~~~or
\label{e1}
\end {equation}
\begin {equation}
\rho(t)=\rho(0)\frac{C(0)}{C} e^{|t-t(0)|/\pi*A2}.
\label{e2}
\end {equation}
Where $\rho(0),C(0),t(0)$ is the dust density, cross-section and angle 
at the beginning of the arc, and $A1, A2$ are the density exponents
to model additional phenomena e.g. dust destruction. The ring was 
located in the orbital plane. In the following we describe the effect
of various free parameters on the light-curve of this object.

{\it Inclination.} 
We carried out a sequence of calculations for 
the most straightforward
assumption that the ring is in the orbital plane of the planet and 
that this orbital plane has an inclination of $i=90$ deg with respect 
to the plane of sky. It turned out that the observed core of the 
transit is too sharp. This indicates that the transit is not entirely
edge on but may have $i<84$ deg. An alternative, explanation, could be 
that the $R^{*}/a$ parameter is overestimated or that the dust 
particle size in the close vicinity of the planet is larger,
 which could narrow 
the transit by filling its ingress and egress by the scattered light 
with sharper forward scattering peak. There is an interesting effect 
that smaller inclinations or grazing transits require slightly higher 
dust densities and that the forward scattering peak gets more 
pronounced. Ultimately, for the non-transiting planet, the light-curve
shows the transit brightening only.
The inclination $i=82$ seems to reproduce the observations well
and, if not mentioned otherwise, we will use this as a default value
in the following calculations. This value corresponds to the impact 
parameter $b=0.60$ expressed relative to the stellar radius or 
$b=0.39 R_{\odot}$ in absolute units.
It is in agreement with the result of Brogi et al. (\cite{brogi12}) 
who estimated several values of the impact parameter in the range 
0.46-0.63.

{\it Geometrical cross-section of the tail.} 
Unfortunately, this quantity cannot be constrained very well. 
There is a strong degeneracy between this quantity,
dust density and optical depth along the line of sight.
It stems from the Eqs. \ref{e1} \ref{e2} and best fitting light-curves 
tend to have the same product $\rho C$. An increase in the cross-section 
by a factor of 2 results in decreasing the dust density by a factor 
of 2 and decreasing the optical depth through the tail by the same 
factor.

\begin{figure}
\centering
\includegraphics[angle=-90,width=8.8cm]{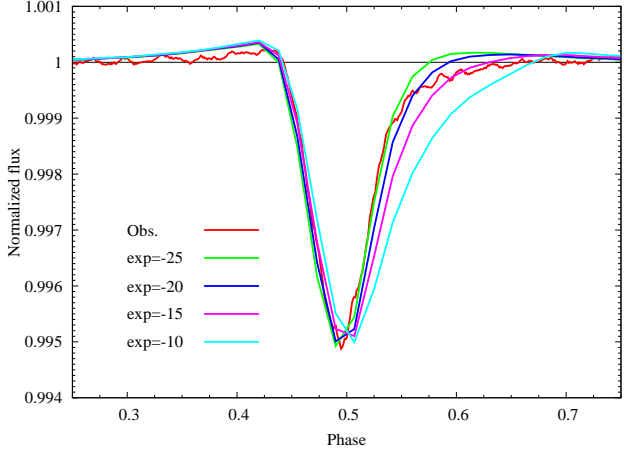}
\caption{
The effect of the dust density profile on the light-curve. 
Higher exponents fit better the core while lower exponents fit better
the tail. In any case, the dust density in the tail rapidly decreases 
with the (angular) distance from the planet which indicates that 
the dust is either quickly destroyed or scattered out of the tail. 
}
\label{fexp}
\end{figure}

{\it Dust density profile.} 
The shape of the transit is highly asymmetric with steep ingress and
long shallow egress. This is the main argument for the idea of a 
comet-like tail and it suggests that the dust density decreases 
steeply along the tail. This density profile is described by the 
Eqs. \ref{e1} and \ref{e2} and is one of the most important free 
parameters. 
This is illustrated in the Figure \ref{fexp} which was calculated 
assuming $i=82$ deg, characteristic particle size of about 
0.1 micron, and iron contaminated pyroxene. The best results 
are achieved when $A1, A2$ are about -20 and it will be our default 
value when studying other effects. 
This is a very steep function of angular distance from the planet 
and it indicates that the dust grains are either effectively destroyed 
or removed from the tail. 
Higher exponents fit the core better while lower exponents fit
the tail better.
This also suggests that the object may consists of two different 
components. Each density exponent requires different density $\rho(0)$ 
to fit the depth of the transit properly and this depends also on 
the particle size and the chemical composition of the dust.
In the Table \ref{t1} we give the dust density $\rho(0)$ and 
the maximum optical depth through the tail in front of the star 
corresponding to a particular density exponent for three kinds of 
species: pyroxene with 20\% of iron, enstatite (pyroxene with
0\% of Fe), and forsterite. 
The calculations assumed an inclination of 82 degrees and 
the geometrical cross-section at the beginning and end of the tail of 
0.01 and 0.09 $R_{\odot}^2$, respectively. 
Notice that the density of the Fe 
contaminated pyroxene and enstatite do not differ considerably
for larger particles but enstatite requires higher densities
for smaller particles. This is because opacity at these wavelengths
for larger particles is dominated by the scattering which is not
affected by the Fe content while opacity of smaller grains
is dominated by the absorption and pyroxenes with higher Fe content
have higher optical opacity. Densities of forsterite do not differ
much from those of enstatite and are only slightly lower for smaller
particles.

\begin{figure}
\centering
\includegraphics[angle=-90,width=8.8cm]{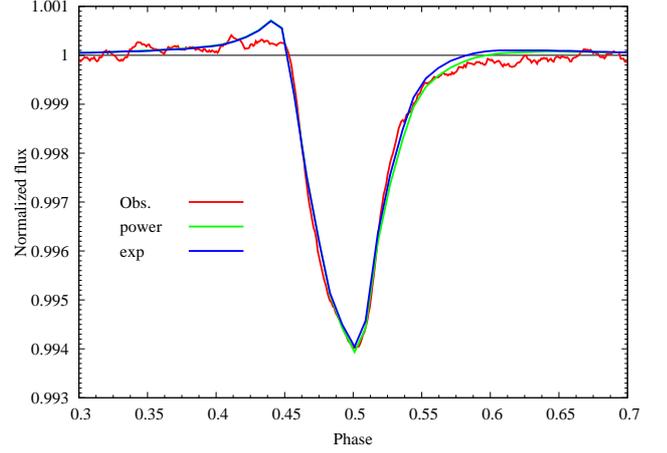}
\caption{
The comparison between the exponential (exp) and power law (power)
density profile.
Both profiles are very similar, exponential profile fits slightly 
better the core while a polynomial profile fits slightly better 
the tail (see the text).
}
\label{fexppow}
\end{figure}

Figure \ref{fexppow} illustrates the difference between the power law
behaviour (Eq.\ref{e1}) and the exponential behaviour (Eq.\ref{e2})
for $A1=A2=-20$, $\rho(0)=5\,10^{-15}$, for an iron contaminated 
pyroxene with 1 micron grains. There is only a little difference
and an exponential profile fits the core slightly better while 
a polynomial profile fits the tail slightly better.

\begin{table}
\caption[]{Density exponent $A2$, dust density $\rho(0)$ at the edge of 
ring $\times 10^{-15}$ g\,cm$^{-3}$, and maximum optical depth $\tau$
for pyroxene with 20\% of iron, enstatite, forsterite and for various 
dust particle radii. Model assumes that the geometrical 
cross-section of the ring is 0.01 and 0.09 $R_{\odot}^2$ at the 
beginning and the end of the ring, respectively.}
\label{t1}
\centering
\begin{tabular}{l|ll|ll|ll}
\hline\hline
& \multicolumn{2}{c}{pyroxene} & \multicolumn{2}{c}{enstatite} 
& \multicolumn{2}{c}{forsterite}\\
\hline
A2   & $\rho$ & $\tau$  & $\rho$ & $\tau$ & $\rho$ & $\tau$\\
\hline
\multicolumn{7}{c}{0.01 micron}\\
-10  &  42  &  0.066 & 110  & 0.064  & 90   & 0.064 \\
-15  &  54  &  0.081 & 150  & 0.082  & 120  & 0.080 \\
-20  &  66  &  0.095 & 178  & 0.094  & 145  & 0.094 \\
-25  &  79  &  0.110 & 215  & 0.11   & 177  & 0.11  \\
\hline 
\multicolumn{7}{c}{0.1 micron}\\
-10  & 0.26 &  0.075 & 0.30 & 0.073  & 0.26 & 0.072 \\
-15  & 0.33 &  0.091 & 0.38 & 0.086  & 0.33 & 0.086 \\
-20  & 0.41 &  0.110 & 0.46 & 0.10   & 0.40 & 0.10  \\
-25  & 0.49 &  0.120 & 0.56 & 0.12   & 0.49 & 0.12  \\
\hline
\multicolumn{7}{c}{1 micron}\\
-10  & 3.0  &  0.14  &  3.0 & 0.13   & 3.0  & 0.13  \\
-15  & 3.8  &  0.17  &  3.8 & 0.15   & 3.8  & 0.15  \\
-20  & 4.6  &  0.19  &  4.6 & 0.18   & 4.6  & 0.18  \\
-25  & 5.6  &  0.23  &  5.6 & 0.21   & 5.6  & 0.21  \\
\hline
\end{tabular}
\end{table}

\begin{figure}
\centering
\includegraphics[angle=-90,width=8.8cm]{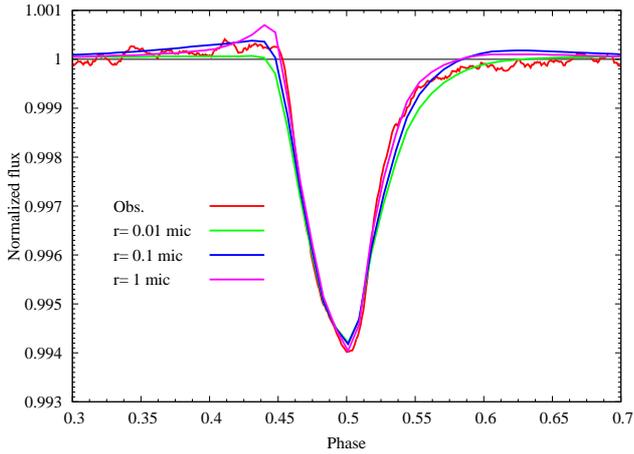}
\caption{
Short cadence observations vs the models.
The effect of the particle size on the light-curve.
There is a strong degeneracy between this parameter and dust density.
However, notice that larger size particles fit better the transit core
whereas smaller size particles fit better an extended egress.
}
\label{size}
\end{figure}

{\it Particle size.}
This parameter is very important since it affects the absorption and
scattering opacity and emissivity which is illustrated in the 
Figure \ref{f2b}. Unfortunately, due to this dependence  there is 
a strong degeneracy between this parameter, dust density and geometrical 
cross-section of the tail. This degeneracy is demonstrated in 
the Figure \ref{size} which depicts a comparison between 
the theoretical light-curves and Kepler short cadence observations. 
It is possible to fit the transit with dust particles of  
different size and different dust densities. 
This degeneracy complicates precise estimates of 
the particle size and dust density from the shape of the transit.
In the Table \ref{t1} we list various possible combinations
of the edge density and density exponent for different particle radii
and different grains.
Notice that 0.1 micron particles require substantially lower dust
densities to reproduce the transit than 1 micron particles which
in turn require much lower densities than 0.01 micron particles.
This can be understood using the Figure \ref{f2b} which shows that
at 600 nm the opacity is dominated by the scattering on 0.1 micron 
particles followed by the scattering on 1 micron particles, and
absorption on 0.01 micron grains.

\begin{figure}
\centering
\includegraphics[angle=-90,width=8.8cm]{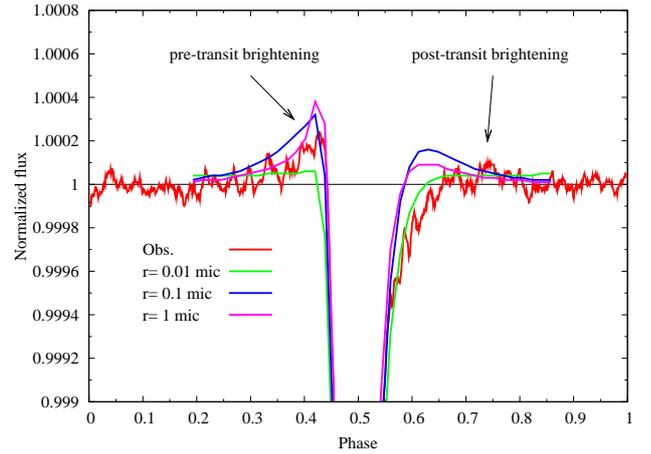}
\caption{
A more detail look on the effect of the particle size on the 
light-curve. Long cadence observations vs models.
Notice a significant pre-transit brightening and less
significant post-transit brightening which are caused by 
the strong forward scattering. It is rather sensitive 
to the particle size. Dust particles with typical radii 
of about 0.1-1 micron fit better the transit core whereas 
0.01-0.1 micron particles fit better an extended egress
and post transit brightening.
}
\label{sizezoom}
\end{figure}

Nevertheless, particle size does leave a few interesting imprints on 
the light-curve. Notice the strong pre-transit brightening and a less
pronounced post-transit brightening which are quite sensitive to 
the particle size.
It is best seen in the long cadence observations since there are
more of them and have less scatter compared to the short cadence data
(see Figure \ref{sizezoom}).
The pre-transit brightening is caused by the strong forward scattering.
This forward scattering is strongly sensitive to the particle size 
as demonstrated in the Figure \ref{f2}. This is why the pre-transit 
brightening is so sensitive to the particle size and can be used 
to estimate the dust particles size.
Larger size particles (0.1-1 micron) or they combination fit better 
the transit core and the pre-transit brightening while smaller size 
particles (0.01-0.1 micron) or their combination fit better the slow 
egress and post-transit brightening.
This is just another indication that the planet is not uniform. 
The particle size may decrease as a function of the angular distance 
from the planet which can be considered as an indication for the dust 
destruction in the tail. Notice that the larger particles exhibit 
a narrower transit than the smaller particles. This is because larger 
particles fill the ingress and egress by the strong forward scattered 
light. This may cause another degeneracy between the particle size,
orbital inclination, and density exponent since they all affect
the width of the transit.
In an independent analysis, Brogi et al. (\cite{brogi12}) 
using different model and analysis estimated the particle size
of about 0.1 micron in the tail. However, in their analysis
they did not use real dust opacities, phase functions were approximated 
by the analytical Henyey-Geenstein phase functions, and the finite
radius of the source of light was ignored.

\begin{figure}
\centering
\includegraphics[angle=-90,width=8.8cm]{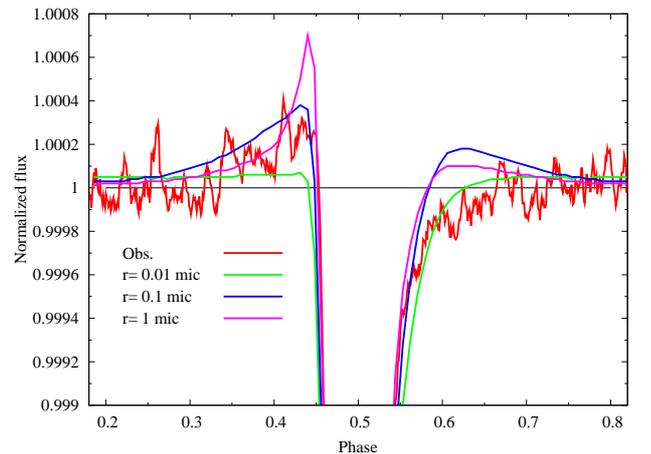}
\caption{
The same as in Figure \ref{sizezoom} but for the Kepler short cadence
observations. The phases 0.3-0.4 are best reproduced by 1 micron
particles, phases at about 0.45 might be affected and screened
by the streams which are not modelled here, phases 0.5-0.55
are best reproduced by the 1 micron particles, phases near 0.55-0.57
are best fit with the 0.1 micron particles whereas later phases further
in the tail are best reproduced with particles that have radii
of about 0.01 micron.
}
\label{sizezoomsc}
\end{figure}

The latest Kepler short cadence observations are also very helpful
in constraining the particle size.
Their comparison with the theoretical light-curves is in 
Figure \ref{sizezoomsc}.
If the calculations are to be interpreted solely in terms of 
the particle size then the phases 0.3-0.4 are best reproduced by 
1 micron particles. Phases at about 0.45 might be affected and screened
by the streams mentioned in the next section which are not 
modelled here. Phases 0.5-0.55 are best reproduced by the 1 micron 
particles. Phases near 0.55-0.57 are best fit with the 0.1 micron 
particles whereas even later phases further in the tail are best 
reproduced with particles that have radii of about 0.01 micron.
Alternative explanation could be that the tail is composed
of several chemical or structural components which have different
A1 or A2 parameters or destruction lifetimes.

{\it Tail morphology.}
The above mentioned ring/arc morphology of the tail seems to
reproduce the observations surprisingly well.
The tail is about 60 degrees long. 
The density in the tail decreases rapidly along the tail
and becomes negligible beyond 60 degrees from the planet.
Apart from the clear pre-transit brightening, notice a weak
post-transit brightening in the Figure \ref{sizezoom}.
This can happen if the tail is terminated by a relatively
sudden drop in the density or is dispersed out of the orbital plane,
reaches the vertical thickness greater than the radius of the star
and there is still a small but non negligible fraction of particles 
with the size of the order 0.1 micron.
Apart from the above mentioned tail morphology, I tried also some
other geometrical models.
For example, a dusty tail of the shape of a cone pointing
at different angles from the planet. Such geometry would produce very 
strong post-transit brightening which would not be compatible with 
the observations.

Theoretical light-curve for 1 micron particles fits the pre-transit
brightening best but shows a slight excess compared to 
the observations. This is best seen in the comparison with 
the short cadence observations in the Figure \ref{sizezoomsc}.
This excess might disappear if a more realistic model of the dust 
morphology is considered.
Notice, that hydrodynamical simulations of Bisikalo et al. 
(\cite{bisikalo13}) predict two streams emanating from the planet.
One leaves the planet via the L2 point and is deflected in the direction
opposite to the orbital movement which might result into the comet
like tail like the one we observe and model here.
The other leaves the planet via the L1 point
and is deflected in the direction of the orbital motion.
This stream is well known in the interacting binaries.
Such stream might wipe out the excess emission in the pre-transit
brightening seen in the theoretical light-curves. 
We do not model this second stream in this paper
since we consider it too premature at this point.

\begin{figure}
\centering
\includegraphics[angle=0,width=8.8cm]{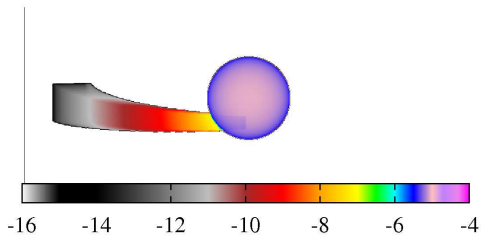}
\caption{
A 2D image (logarithm of intensity) of the planet with a comet-like 
tail during the transit.
One can clearly see the tail. Its intensity rapidly decreases 
due to decreasing dust density and strong forward scattering.
Notice also the optically thin absorption on the limb darkened
stellar disk. Calculated for 1 micron pyroxene grains.
}
\label{2d}
\end{figure}

Our model of the planet and what happens during the transit is 
illustrated in the Figure \ref{2d}. This is a 2D image of 
the model which shows the logarithm of the intensity.
One can clearly see the tail with a very steep intensity gradient
which becomes much brighter close to the star due to the forward 
scattering and density gradient. Notice also the optically thin 
absorption on the stellar disk.

{\it KIC 12557548b in the more general context.}
It is very interesting to note that this pre-transit brightening is
analogous to the mid-eclipse brightening (MEB) observed in 
some long period eclipsing binaries like
$\epsilon$ Aur (Budaj \cite{budaj11a}) or AZ Cas (Galan et al.
\cite{galan12}).
MEB in $\epsilon$ Aur may be due to dark flared dusty disk 
eclipsing F super-giant and forward scattering plays an important role.   
MEB in AZ Cas may also be caused by the forward scattering during
the primary eclipse. During this eclipse the hotter B star 
is eclipsed by the cool M super-giant. Dust surrounding the cool
component can scatter the light from the hidden hot star to 
the observer. Such effect was predicted and calculated
in Section 2 of Budaj \cite{budaj11a}.
Since very similar dust properties and physics can explains 
the light-curves of all three objects it gives stronger footing to our 
explanation of the shallow MEB in $\epsilon$ Aur.
This model is in contrast with the geometrical model of the inclined 
disk and the star peeking through the hole.
In spite of the fact that these are very different objects it looks like
they are subject to similar physical processes, can be modelled with 
similar tools, and the knowledge obtained in one, can be applied 
to the other. This analogy suggests that there might also be a tail 
of dust emanating from the disk of  $\epsilon$ Aur which could explain 
non-symmetric behaviour of gas and dust absorption during the eclipse 
with egress absorption being stronger and longer 
(Martin et al. \cite{martin13}, Leadbeater et al. \cite{leadbeater13}). 
On the other hand there might be a dusty disk formed inside the Hill
sphere of KIC12557548b which would feed into the comet-like 
tail.
 
There is one more thing which needs to be mentioned.
Long cadence observations have a relatively long exposure (30 min)
which might smear any potential sharp features in the light curve. 
That is why the theoretical light-curves were convolved with 
the box-car which had a width corresponding to the exposure time
(0.031 phase units).
Short cadence observations have exposure time (1min) which is much
shorter than the box-car smoothing applied to the observed data.
Consequently, theoretical light-curves used for comparison with 
the short cadence data were smoothed with the box-car which had a
width corresponding to the box-car width applied to the observations
(0.01 in phase units). This had very little effect on 
the result.

\section{Conclusions}

The light-curve of this planet candidate was reanalysed using first 
14 quarters of the Kepler data including first short cadence 
observations. 

Orbital period of the planet was improved.
We searched for the long term period changes but found no convincing
evidence of such changes.

Quasi periodic variability in the tail of the planet with the period 
of about 1.5 year was discovered.

We modelled the light-curve of KIC012557548 using SHELLSPEC code 
with the following assumptions: spherical dust grains,
different dust species (pyroxene, enstatite, forsterite),
different particle sizes, Mie absorption and scattering, finite radius 
of the source of light (star) and that the medium is optically thin. 
We proved that its peculiar light-curve is in agreement with the idea 
of a planet with a comet like tail. 
A model with a dusty ring in the orbital plane which has an inclination
of about 82 degrees, exponential or power law density profile 
($A1=A2=-20$) fits the observations surprisingly well.

We confirmed that the light-curve has a prominent 
pre-transit brightening. There is an indication of a less prominent 
brightening after the transit, both are caused by the forward 
scattering.

Dust density in the tail is a steep decreasing function of the distance
from the planet which indicates significant destruction of the tail
caused by the star.

Transit depth is a highly degenerate function of the particle size,
dust density, and other dust properties. Various combinations of 
them were estimated and tabulated.
There will also be a degeneracy between the inclination, particle size,
and the density exponent.

However, the forward scattering and the pre(post)-transit brightening 
are quite sensitive to the particle size.
Consequently, we estimated the particle size (radius) of the grains in 
the head of the tail from the pre-transit brightening to be about
0.1-1 micron. There is an indication that the particle size is larger
at the head and decreases along the tail to about 0.01-0.1 micron.

We argue that there are several indications that the 'planet' is not 
homogeneous and that it consists of several components.
The component that is responsible for the transit core and the other
responsible for the tail. Components may have different grains with 
different density profiles, and/or particle size.

It is interesting to note that this planet's light-curve with 
pre-transit brightening is analogous to light-curves of some 
interacting binaries with mid-eclipse brightening, particularly 
$\epsilon$ Aur and forward scattering plays an important role 
in all of them.

\begin{acknowledgements}
I thank Dr. M. Kocifaj for his help with his code, Dr. R. Komzik
and Prof. R. Stellingwerf for their help with the PDM code, 
Prof. J. Schneider, Dr. L. Hambalek, and T. Krejcova for their 
comments and discussions.
This work was supported by the VEGA grants of the Slovak Academy of
sciences Nos.  2/0094/11, 2/0038/13, by the Slovak Research
and Development agency under the contract No.APVV-0158-11, and
by the realization of the
Project ITMS No. 26220120029, based on the supporting operational 
Research and development program financed from the European Regional 
Development Fund.

\end{acknowledgements}

\end{document}